# Qochas on Andean highlands

Amelia Carolina Sparavigna
Dipartimento di Fisica
Politecnico di Torino, Torino, Italy

On the Andean highlands, the "qochas" are lakes or ponds of natural or artificial origin. An ancient agricultural technique is based on their use. Linked together by a network of canals, qochas form a system of water and soil management, alternately used for crops or pasture. The concave structure of qochas controls the strong evaporation produced by solar radiation and wind blowing. Qochas can be observed in the satellite imagery of Google Maps.

In Quechua, the term "qocha" refers to a lake, pool or pond of natural or artificial origin. A monumental qocha of the Inca period is in the outskirts of Cuzco. Near the Rodadero, a rocky hill with numerous stairwells and benches carved into the stone, a spring is providing water to the Qocha Chincanas, a round artificial pond for ceremonial purposes (Qocha Chincanas is show in the upper panel of Fig.1). Qocha is the name of some pottery too.
An ancient agricultural technique is based on the use of qochas. They appear at the vertices of a network of canals. For the agricultural purposes, the qochas are the depressions, natural or dug by men on the ground surface, that can be flooded. These qochas are found in some areas on the Andean highlands of Peru and Bolivia, near the Titicaca Lake, at an average altitude of 4000 meters. They are very numerous and dense in the department of Puno (Peru), where they were discovered and documented.
A beautiful book entitled "Agricultures Singulières", which is an "around the world" in several discussions on ancient agricultural techniques, devotes a chapter to the Qochas of Andean regions. According to [1], the qochas are linked together to form a real hydraulic system: the term is then referring both to a single pool and to the whole system. The land among them is used for breeding animals. The qochas form one of the many systems for the management of water and soil that allowed the ancient Andean people to farm and survive. The qochas are arranged in some basic shapes, squares and circles. The most common shape is round and measures 30 meters to 200 meters in diameter. The qocha serves alternatively for crops and pasture and also as water tank. It grows mainly potatoes, cañihua, quinoa and other crops on rotation. The structures gives potatoes the first year, quinoa, oats or barley the second year and then it rests for a year as a fallow pasture [1]. The system allows a flexible control of the water of rain in a region that faces a succession of severe droughts followed by floods.
In his article [2], Pierre Morlon is discussing the physical properties of qochas with circular structures. The rain is collected in the pond by radial and circular canals among earthworks (see Fig.2). The concave structure is reducing the strong evaporation provoked by the solar radiation and the wind blowing. Fig.2 is showing one of these concave structures. Of the original shape, only one half is persisting. I have seen this qocha in 2010 and recently in 2011 by means of Google Maps. The image downloaded in 2010 shows the structure with wet canals. Adriano Forgione suggested me that such a structure could give the possibility to measure time [3]. In fact, the concave radial structure could provide several reference points to the farmer, to check by means of dawns and sunsets the proper seed-time.

Figure 3 is another image obtained by Google Maps, showing several qochas. Two of them display radial and circular earthworks. A project on anthropological studies of the Pennsylvania State University, the AnthSpace Analytical Cartography Lab, is monitoring the qochas in Peru by means of satellites [4]. As in the case of the raised fields, the qochas can be seen in the Google Maps imagery. The details of these structures can be enhanced with several tools of image processing [5-7].

Ref.2 is telling that up to fifty years ago, waru-warus and qochas were much more extensive on the land but they were deliberately destroyed. In fact, thirty years ago, several projects for the restoration of waru-warus and qochas have been supported [8]. Waru-warus are the other agricultural structures which can be seen near the Titicaca Lake. They are earthworks, also known as "raised fields" (see Figure 4), five to ten meters large and quite long, even several hundred meters, creating an extensive network of canals.

Waru-warus are the remains of a huge agricultural system [9-14]. According to Ref.15, the structure of waru-warus contributes to frost mitigation during the growing season. The paper uses a two-layers model to explain the role played by the canals in the nocturnal heat dynamics and the cold mitigation process. The model shows that the heat flux emanating from the canals and the water condensation on the crop both contribute to the mitigation effect. A similar study and modelling could be interesting to investigate the thermodynamics of qochas.

Besides qochas and waru-warus, other ancient structures for the agriculture can be found in Peru and Bolivia. There are the "bofedales", artificial wetlands, the "andenes", which are terraced hills and the "puquios". The puquios of Nasca are a system of subterranean filtration galleries that provide water for irrigation and domestic uses in the middle portions of the Nasca and in other areas of Peru [16].

According to [1], the system of qochas of the altiplano around Lake Titicaca is the least studied and therefore we find a few references about it. Only in 1962 these qochas were mentioned for the first time by the archaeology students working in Puno. The structures have a pre-Incaic origin as shown by the fragments of pottery nearby found. Although the system is probably earlier, it is associated with the Pukara culture.

Pukara was a major center for hundreds of years since 1300 BC. Between 250 BC and 380 AD, it became an important religious site, densely populated. Pukara had a highly structured society able to provide a centralized management of the water. Some authors suggest that the development and maintenance of agricultural structures, including qochas were equally well implemented by local community groups [9-11]. Even assuming that farmers have used existing depressions in which the water of rain accumulates, it is certain that a large workforce and a long period of time were needed to develop this complex system over a such large area. Then the area was gradually abandoned with the rise of Tiwanaku, with which Pukara had trade and social links. All the archaeological and satellite evidences show that the Andean civilization, which extended its influence to the northern Chile, practiced a florid agriculture based on the construction of terraces on hills, raised fields and qochas. After the fall of Tiwanaku, about 1000 AD, the site of Pukara was repopulated and qochas used again [1]. The region endured a long period of drought that could make the use of qochas essential for the survival of local populations. Later, the Huari and Inca developed the andenes system, but qochas continued to be cultivated. The Spanish conquest led to a severe local depopulation which produced a partial ruins of all farming structure, but the qochas were still used. The reason is that these structures are quite simple and small and can be managed by a single family. Qochas are then appearing as those structures that primarily thrive when environmental conditions (natural or social) are undermining agriculture and local communities [1].

Today whole sectors of qochas have been abandoned or deteriorated for reasons of soil salinity and because of the agricultural mechanisation in the areas of haciendas. As told in [1], more than 20,000 cups are still used by the Andean people. Their persistence in the history is telling that their use is a more sustainable technique for the agriculture in Andean environment.

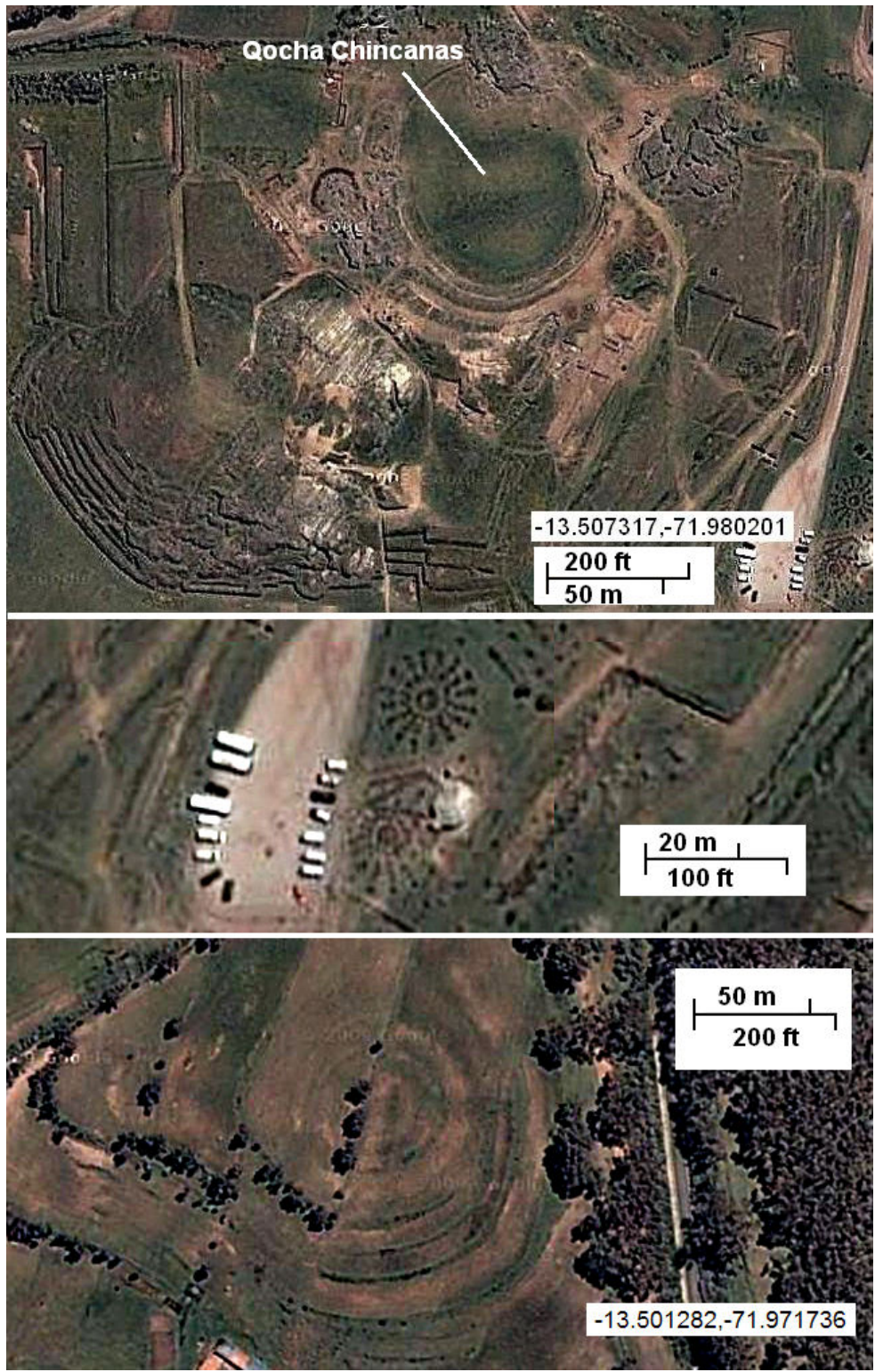

**Figure 1:** Near the Rodadero of Cuzco, a spring is providing the water to Qocha Chincanas, a round artificial pond for ceremonial purposes (upper panel). In the middle, the image shows the detail of two small qochas with a radial structure near the parking. In the lower panel, a round structure, probably another large qocha near Rodadero.

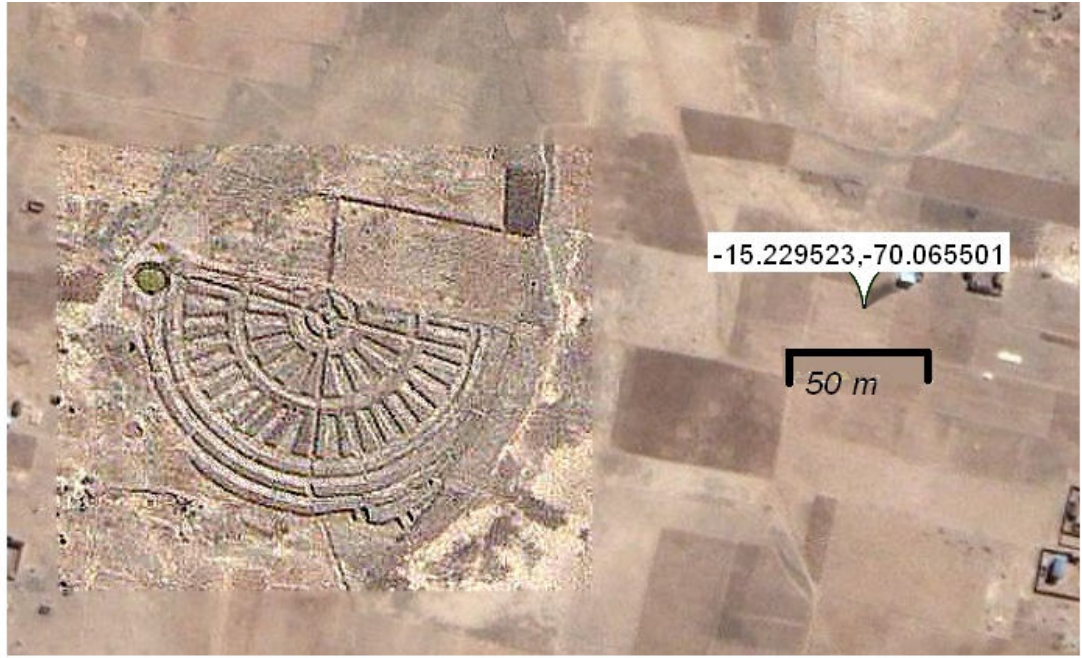
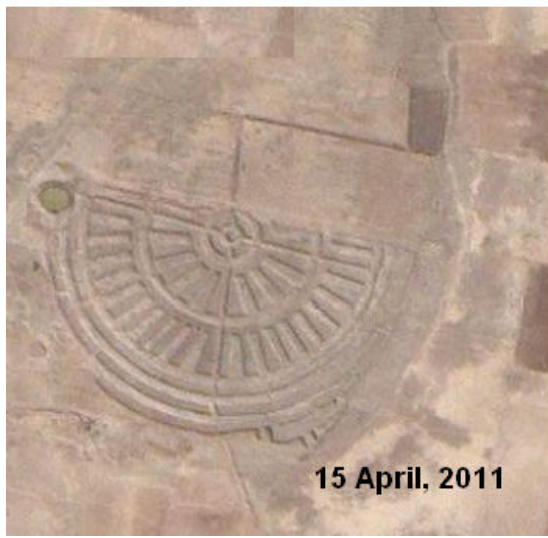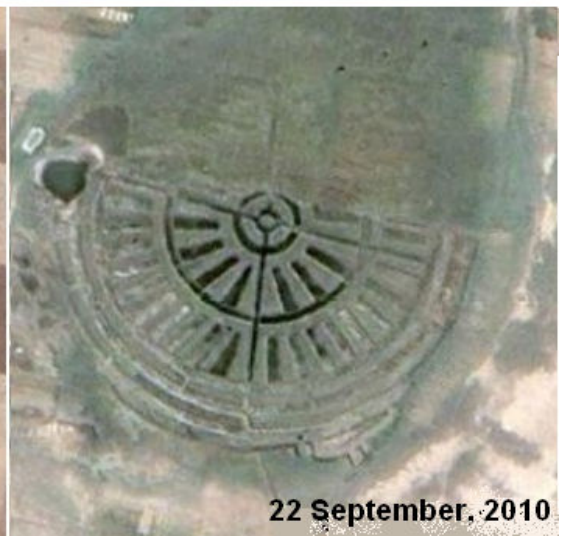

**Figure 2**: Some qochas have circular structures. The rain is collected by radial and circular canals among earthworks. The concave structure is reducing the strong evaporation provoked by the solar radiation and the wind blowing. Of the original shape, only one half is persisting. The image downloaded in 2010 shows the structure with wet canals.

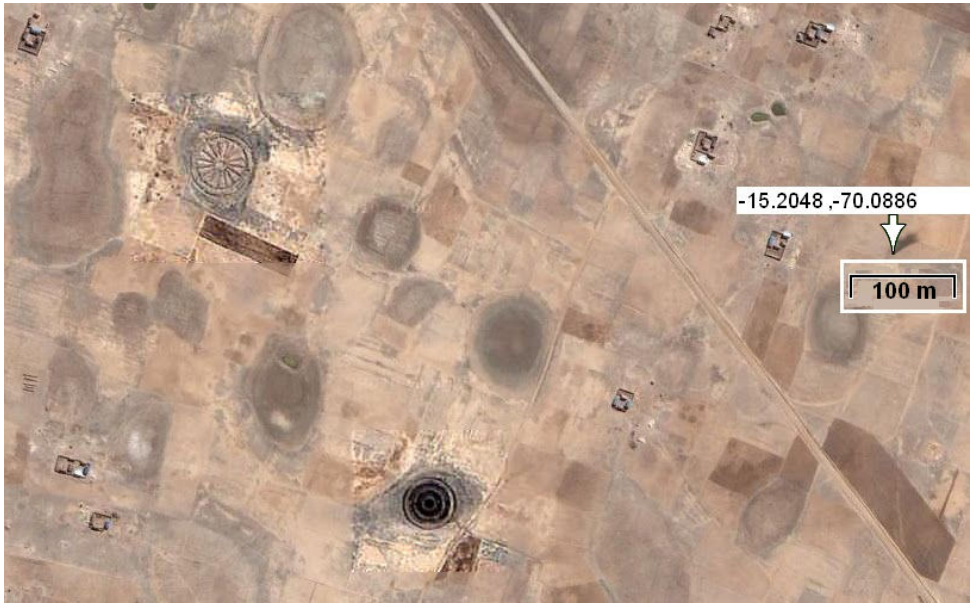

**Figure 3**: In this image obtained by means of the Google Imagery, we can see several qochas. Two of them display very well the radial and circular earthworks.

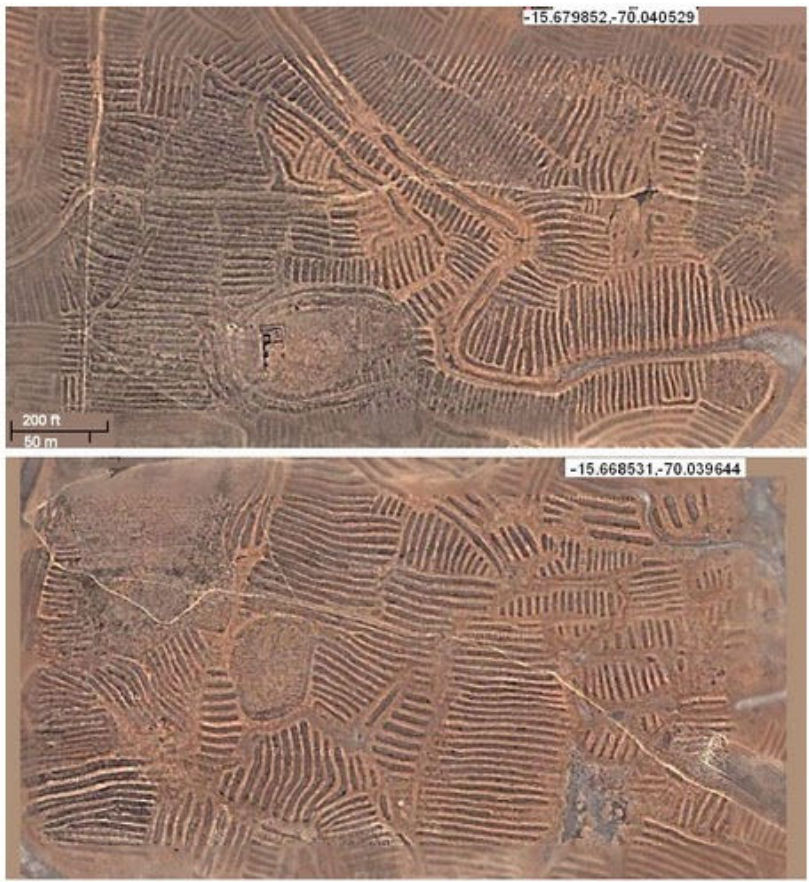

**Figure 4**: The network of raised fields near Titicaca Lake.